\documentclass[aps,showpacs,twocolumn,nofootinbib]{revtex4}
\usepackage{amssymb}
\usepackage{natbib}
\usepackage{amsmath}
\usepackage{amsfonts}
\usepackage{graphicx}
\usepackage{mathrsfs}
\usepackage{dcolumn}
\usepackage{bm}
\usepackage{color}
\definecolor{rot}{rgb}{0.75,0.05,0.25}
\definecolor{hellgrau}{gray}{0.5}
\definecolor{blau}{rgb}{0,0,0.7}

\def\Tr{\mbox{Tr}}

\begin{document}

\title{Construction of microcanonical entropy on thermodynamic pillars}

\author{Michele Campisi}
\email{michele.campisi@sns.it}
\affiliation{NEST, Scuola Normale Superiore \& Istituto Nanoscienze-CNR, I-56126 Pisa, Italy}
\date{\today }
\begin{abstract}
A question that is currently highly debated is whether the microcanonical entropy should be expressed as the logarithm of the phase volume (volume entropy, also known as the Gibbs entropy) or as the logarithm of the density of states (surface entropy, also known as the Boltzmann entropy).
Rather than postulating them and investigating the consequence of each definition, as is customary, here we adopt a bottom-up approach and construct the entropy expression within the microcanonical formalism upon two fundamental thermodynamic pillars: (i) The second law of thermodynamics as formulated for quasi-static processes: $\delta Q/T$ is an exact differential, and (ii) the law of ideal gases: $PV=k_B NT$. The first pillar implies that entropy must be some function of the phase volume $\Omega$. The second pillar singles out the  logarithmic function among all possible functions. Hence the construction leads uniquely to the expression $S= k_B \ln \Omega$, that is the volume entropy. 
As a consequence any entropy expression other than that of Gibbs, e.g., the  Boltzmann entropy, can lead to inconsistencies with the two thermodynamic pillars. We illustrate this with the prototypical example of a macroscopic collection of non-interacting spins in a magnetic field, and show that the Boltzmann entropy severely fails to predict the magnetization, even in the thermodynamic limit. The uniqueness of the Gibbs entropy, as well as the demonstrated potential harm of the Boltzmann entropy, provide compelling reasons for discarding the latter at once.
\end{abstract}
\pacs{
} 
 \maketitle

\section{Introduction}
The recent paper by Dunkel and Hilbert titled ``Consistent thermostatistics forbids negative absolute temperatures'' \cite{Dunkel14NATPHYS10} has triggered a vigorous debate on whether the Boltzmann entropy (alias the surface entropy, Eq. \ref{eq:Somega}) or the Gibbs entropy (alias the volume entropy, Eq. \ref{eq:SOmega}) is the more appropriate expression for the thermodynamic entropy of thermally isolated mechanical systems 
\cite{Sokolov14NATPHYS10,Vilar14JCP140,Frenkel15AJP83,Dunkel14ArXiv1403.6058,Schneider14arXiv1407.4127,Dunkel14arXiv1408.5392,Hilbert14PRE90,Swendsen14arXiv1410.4619}. 
The thermodynamic consistency of the Gibbs entropy has been a leitmotiv that sporadically recurred in the classical statistical mechanics literature. It started with Helmholtz \cite{Helmholtz95INBOOK}, Boltzmann \cite{Boltzmann84CJ98}, and Gibbs \cite{GibbsBook}, it continued with P. Hertz \cite{Hertz10AP338a} Einstein \cite{Einstein11AP34}, and others \cite{Schlueter48ZNA3,Munster69book}, until it has been reprised recently by various authors \cite{Berdichevsky91PRA43,Pearson85PRA32,Adib04JSP117,Campisi05SHPMP36,Dunkel06PHYSA370}.
This line of research culminated with the work of Ref. \cite{Hilbert14PRE90}, showing that the Gibbs entropy complies with all known thermodynamic laws and unveiling the mistakes apparently incurred into the arguments of its opponents \cite{Vilar14JCP140,Frenkel15AJP83,Schneider14arXiv1407.4127,Swendsen14arXiv1410.4619}. 

While the work of Ref. \cite{Hilbert14PRE90} is characterised by a top-down approach (namely, one postulates an entropy expression and then investigates compliance with the thermodynamic laws) here we adopt instead a bottom-up approach:
we begin from the thermodynamic laws and construct the expression of the microcanonical entropy on them. In particular we base our construction on the following two fundamental pillars of thermodynamics. 1) The second law of thermodynamics as formulated by Clausius for quasi-static processes, namely, $\delta Q/ T = dS$, which says that $1/T$ is an integrating factor for $\delta Q$, and identifies the entropy with the associated primitive function $S$. 2) The equation of state of an ideal gas $PV=k_BNT$. 

Our construction, based on the mathematics of differential forms, leads uniquely to the Gibbs entropy; see Sec. \ref{sec:construction}.
As a consequence the adoption of any expression of entropy other than the Gibbs entropy, e.g., the Boltzmann entropy, may lead to inconsistency with the fundamental pillars. This will be illustrated with a macroscopic collection of spins in a magnetic field. As we will see the Boltzmann entropy severely fails to predict the correct value of the magnetization, and even predicts a nonexistent phase transition in the thermodynamic limit, see Sec. \ref{subsec:failure}. This provides a compelling reason for discarding the Boltzmann entropy at once.

The present work thus complements the work of Ref. \cite{Hilbert14PRE90} by stating not only the compliance of the Gibbs entropy with the thermodynamic laws, but also its \emph{necessity} and \emph{uniqueness}: thermodynamic entropy has to be expressed by means of Gibbs formula and no other expression is admissible. 

Together with Ref. \cite{Hilbert14PRE90} the present work appears to settle the debated issue.

\section{Definitions}
We recall the definitions of Boltzmann and Gibbs entropies within the microcanonical formalism \cite{Campisi05SHPMP36}:
\begin{align}
S_B(E,\bm{\lambda})  &= k_B \ln \left [\omega(E,\bm{\lambda})\varepsilon \right ]\label{eq:Somega}\, , \\ 
S_G(E,\bm{\lambda})  &= k_B \ln \Omega(E,\bm{\lambda}) \label{eq:SOmega} \, ,
\end{align}
where 
\begin{align}
\Omega(E,\bm{\lambda}) = \Tr\, \Theta[E-H(\bm{\xi};\bm{\lambda})]
\label{eq:Phi}
\end{align}
denotes the volume of the region of the phase space of the system with energy not above $E$. The symbol $\varepsilon$ stand for some arbitrary constant with units of energy. Here $H(\bm{\xi};\bm{\lambda})$, denotes the Hamilton function of either a classical or a quantum system with degrees of freedom $\bm{\xi}$ and $\bm{\lambda}=(\lambda_1, \lambda_2 \dots \lambda_L)$ denotes external parameters, e.g. the volume of a vessel containing the system or the value of an applied magnetic or electric field \cite{Hilbert14PRE90}. $L$ is their number. In the case of continuous classical systems the symbol $\Tr$ stands for an integral over the phase space normalized by the appropriate power of Planck's constant and possible symmetry factors. For classical discrete systems, $\Tr$ denotes a sum over the discrete state space. For quantum systems $\Tr$ is the trace over the Hilbert space.
The symbol $\Theta$ stands for the Heaviside step function. The symbol $\omega(E,\bm{\lambda})$ stands for the density of states, namely the derivative of $\Omega(E,\bm{\lambda})$ with respect to $E$:
\begin{align}
\omega(E,\bm{\lambda}) = \Tr\, \delta[E-H(\bm{\xi};\bm{\lambda})] = \frac{\partial \Omega(E,\bm{\lambda}) }{\partial E}\, .
\label{eq:Omega}
\end{align}
Here it is assumed that the spectrum is so dense that the density of states can be considered a smooth function of $E$.

\section{The construction}
\label{sec:construction}
The main objective is to link thermodynamic observables, i.e. the forces $F_i$ and  temperature $T$, to the quantities which naturally pertain to both the mechanical Hamiltonian description and the thermodynamic description, i.e., the energy $E$ and the external parameters $\bm{\lambda}$.
As we will see the entropy, $S$ will follow automatically and uniquely once the $F_i$'s and $T$ are linked.

We begin with the thermodynamic forces $F_i$, whose expression is universally agreed upon
\cite{LandauBook5}:
\begin{align}
F_i(E,\bm{\lambda}) = -\left \langle \frac{\partial H}{\partial \lambda_i}\right \rangle \, ,
\label{eq:F}
\end{align}
with $\langle \cdot \rangle$ denoting the ensemble average.
Within the microcanonical framework these are expressed as:
\begin{align}
F_i(E,\bm{\lambda}) = -\Tr\, \left( \frac{\partial H(\bm{\xi};\bm{\lambda})}{\partial \lambda_i} \frac{ \delta[E-H(\bm{\xi};\bm{\lambda})]}{  \omega(E,\bm{\lambda})} \right) \, .
\label{eq:Fmicro}
\end{align}

With the expression of $F_i(E,\bm{\lambda})$ we can construct the differential form representing heat:
\begin{align}
\delta Q = dE + \sum _i F_i(E,\bm{\lambda}) d \lambda_i \, .
\label{eq:deltaQ}
\end{align}
$\delta Q$ is a differential form in the $1+L$ dimensional space $(E,\bm{\lambda})$. It is easy to see that, in general $\delta Q$ is not an exact differential; see, e.g., Ref. \cite{Campisi10AJP78}. 

Before we proceed it is important to explain the meaning of $Q$ within the microcanonical formalism. The idea behind the microcanonical ensemble is that $E$ and $\bm{\lambda}$ are controllable parameters.\footnote{Similarly $\beta$ and $\bm{\lambda}$ are controllable parameters in the canonical formalism} Accordingly, if the system is on an energy surface identified by  $(E,\bm{\lambda})$, the idea is that the experimentalist is able to steer it onto a nearby energy shell $(E+dE,\bm{\lambda}+d\bm{\lambda})$. In practice this is can be a difficult task. It can be accomplished, in principle, in the following way: the experimentalist should first change the parameters by $d \bm{\lambda}$ in a quasi-static way.
This induces a well defined energy change $\delta w = - \sum _i F_i(E,\bm{\lambda}) d \mathbf{\lambda}_i$, which is the work done on the system. This brings the system to the energy shell $(E+\delta w,\bm{\lambda}+d\bm{\lambda})$. To bring the system to the target shell $(E+dE,\bm{\lambda}+d\bm{\lambda})$ the experimentalist must now provide the energy $ d E - \delta w$ by other means while keeping the $\bm{\lambda}$ fixed. For example she can shine  targeted amounts of light on the system, from a light source. After the energy $ d E - \delta w$ is absorbed  by the system (or emitted, depending on its sign), no other interaction occurs and the system continues undisturbed to explore the target shell $(E+dE,\bm{\lambda}+d\bm{\lambda})$. In this framework the light source acts as a reservoir of energy, and the quantity  $\delta Q= d E - \delta w$, identified as heat, represents the energy it exchanges.

According to the second law of thermodynamics in the formulation given by Clausius the inverse temperature $1/T$ is an integrating factor for $\delta Q$. This fundamental statement is often called the \emph{heat theorem} \cite{GallavottiBook}. We recall that an integrating factor is a function $\beta(E,\bm{\lambda})$ such that $\beta\delta Q$ equals the total differential $df$ of some function $f(E,\bm{\lambda})$ called the associated primitive, or in brief, just the primitive. Primitives are determined up to an unimportant constant, which we will disregard in the following. Entropy is defined in thermodynamics as the primitive associated with Clausius's integrating factor $1/T$ \cite{Fermi56Book}: 
\begin{align}
 dS \doteq \delta Q /T \, .
\end{align}
In searching for thermodynamically consistent expressions of temperature within the microcanonical formalism, one should therefore look among the integrating factors of the microcanonically calculated heat differential in (\ref{eq:deltaQ}).
It must be remarked that it is not obvious that one integrating factor exists, because the existence of integrating factors is not guaranteed in spaces of dimensions higher than $2$.  So the existence of a mechanical expression for thermodynamic temperature (hence of the entropy) is likewise not obvious.

It turns out however that an integrating factor  for the differential in (\ref{eq:deltaQ}) always exists.
Finding it is straightforward if one re-writes the forces in the following equivalent form
\begin{align}
F_i(E,\bm{\lambda}) &= -\Tr\, \left( \frac{\partial H(\bm{\xi};\bm{\lambda})}{\partial \lambda_i} \frac{ \delta[E-H(\bm{\xi};\bm{\lambda})]}{  \omega(E,\bm{\lambda})} \right) \nonumber\\
&=  \frac{1}{\omega(E,\bm{\lambda})} \Tr\, \left( \frac{\partial \Theta[E-H(\bm{\xi};\bm{\lambda})]}{\partial \lambda_i}\right) \nonumber \\
&= \frac{1}{\omega(E,\bm{\lambda})} \frac{\partial \Omega(E,\bm{\lambda})}{\partial \lambda_i}\, .
\label{eq:Fomega}
\end{align}
This follows from the fact that Dirac's delta is the derivative of Heaviside's step function. With this, Eq. (\ref{eq:deltaQ}) reads
\begin{align}
\delta Q = dE +  \frac{1}{\omega} \sum_i  \frac{\partial \Omega}{\partial \lambda_i}d\lambda_i \, .
\end{align}
It is now evident that $\omega$ is an integrating factor:
\begin{align}
\omega \delta Q= \omega dE +  \sum_i \frac{\partial \Omega}{\partial \lambda_i}d\lambda_i
= \frac{\partial \Omega}{\partial E}dE +\sum_i   \frac{\partial \Omega}{\partial \lambda_i}d\lambda_i = d\Omega\, ,
\end{align}
$\Omega$ being the associated primitive. This does not mean that $1/\omega$ should be identified with temperature and accordingly $\Omega$ with entropy. In fact if an integrating factor exists, this identifies a whole family of infinitely many integrating factors.

To find the family of integrating factors, consider any differentiable function $g(\Omega)$ with non null derivative $g'$. Its total differential reads:
\begin{align}
d g =  g'(\Omega) d\Omega = [g'(\Omega)\omega]  \delta Q\, .
\label{eq:dg}
\end{align}
This means that any function $\beta(E,\bm{\lambda})$ of the form 
\begin{align}
\beta = g'(\Omega)\omega= \frac{\partial}{\partial E} g (\Omega) \, ,
\label{eq:beta}
\end{align}
is an integrating factor for the heat differential $\delta Q$, and $g(\Omega)$ is the associated primitive. In fact all integrating factors must be of the form in Eq. (\ref{eq:beta}), which is equivalent to saying that all associated primitives must be of the form
\begin{align}
f(E,\bm{\lambda})= g(\Omega(E,\bm{\lambda})) \, .
\label{eq:g(Phi)}
\end{align}
To prove that all primitives must be of the form in Eq. (\ref{eq:g(Phi)}) we consider the adiabatic manifolds, namely the $L$ dimensional manifolds in the space 
$(E,\bm{\lambda})$ identified by the condition that $\Omega = \text{const}$, i.e., $d \Omega = \omega \delta Q =0$. Note that the density of states is a strictly positive function $\omega=\partial \Omega /\partial E>0$. This is because increasing the energy results in a strictly larger enclosed volume in the phase space. Thus, the adiabatic manifolds are characterised by the condition $\delta Q =0$, (i.e., any path occurring on them involves no heat exchanges), and each value of $\Omega$ identifies one and only one adiabatic manifold.
Any  primitive $f(E,\bm{\lambda})$ associated with an integrating factor $\beta$ stays constant on the adiabatic manifolds: $\delta Q = 0\implies \beta \delta Q = 0 \implies df = 0  $ unless $\beta$ diverges, which we exclude here. Hence the only way by which any primitives $f(E,\bm{\lambda})$ and $\Omega(E,\bm{\lambda})$ can both be constant on all adiabatic manifolds is that $f$ is a function of $\Omega$, as anticipated.

Note that this rules out automatically the surface entropy $S_B=k_B \ln[\omega\epsilon]$ because, in general, the density of states cannot be written as a function of the phase volume $\Omega(E,\bm{\lambda})$. 
This is clear for example in the case of an ideal monoatomic gas in a vessel of volume $V$, for which $\Omega(E,V)= \text{const} \times E^{3N/2}V^N$ and  $\omega= (3N/2E)\Omega$ \cite{Khinchin49Book}; see below.

Our derivation above tells us that the second law requires that the entropy, which is one of the primitives, has to be a function $g(\Omega)$ of the phase volume, but does not tell us which function that is. For that we need to identify which, among the infinitely many integrating factors, $\beta = \partial g(\Omega)/\partial E$ corresponds to Clausius's notion of temperature.
We remark that once the function $g$ is chosen, it has to be one and the same for all systems. This is because by adjusting the external parameters $\bm{\lambda}$, whose number and physical meaning is completely unspecified, one can transform any Hamiltonian into any other. This fact reflects the very essence of Clausius's heat theorem, namely, that there exists a unique and universal scale of temperature which is one and the same for all systems \cite{Weiss06AJP74}.

We proceed then to single out the function $g$ that is consistent with the notion of temperature of an ideal monoatomic gas in a vessel of volume $V$, taking its equation of state 
 $PV = k_B NT $ as the definition. The Hamilton function of an ideal monoatomic gas reads
 \begin{align}
H(\bm{q},\bm{p};V) = \sum_{i=1}^{3N} p^2/2m + \phi_\text{box}(\bm{q},V)\, ,
\end{align}
with  $ \phi_\text{box}(\bm{q},V)$ representing the box potential confining the gas within the volume $V$.
 The phase volume  reads \cite{Khinchin49Book}
 \begin{align}
 \Omega(E, V)=  \text{const} \times E^{3N/2}V^N\qquad \text{(ideal gas)}\, .
 \end{align}
Hence, using Eq. (\ref{eq:Fomega}), we obtain for the pressure, $P=-\langle \partial_V H\rangle $:
$P =2E/3V$. Confronting this with the ideal gas law we obtain 
\begin{align}
k_B T = 2E/3N \qquad \text{(ideal gas)}\, ,
\end{align}
consistently with what is known from thermodynamics.
Since 
\begin{align}
\omega= \partial \Omega /\partial E = (3N/2E)\Omega \qquad \text{(ideal gas)}\, ,
\end{align}
 in this case, we readily recognize that $1/T = k_B \omega/\Omega$, namely,
 \begin{align}
\frac{1}{ T(E,V)} = \frac{ \partial}{\partial E} ( k_B  \ln \Omega )\, .
\label{eq:PartialLogPhi}
\end{align}
That is  $g(x) =k_B \ln x$, which singles out the Gibbs entropy,
\begin{align}
S(E,\bm{\lambda}) = k_B \ln \Omega (E,\bm{\lambda}) \, ,
\end{align}
as the primitive associated with the integrating factor corresponding to the thermodynamic absolute temperature
\cite{Hilbert14PRE90}.

In sum: if one accepts the microcanonical expression (\ref{eq:F}) of the forces, the Gibbs entropy represents the \emph{only} expression of 
thermodynamic entropy which is consistent with the second law of thermodynamics, Eq. (\ref{eq:deltaQ}), and the equation of state of the ideal gas. 

\section{Discussion}
\subsection{Ensemble inequivalence}
\label{subsec:inequivalence}

As mentioned above the density of states is definite positive  $\omega > 0$, also, by definition, the volume $\Omega$
 is non-negative. Hence their ratio $k_B T=\Omega/\omega$ is non-negative.
This means that, within the \emph{microcanonical} formalism, negative temperatures are inadmissible. Often the present microcanonical scenario is confused with the more common canonical scenario, where the system stays in a canonical state at all times during a transformation, e.g., Ref. \cite{Schneider14arXiv1407.4127}. This is unfortunate because, as we see below, microcanonical and canonical descriptions are not equivalent for those finite spectrum systems usually discussed in this context. 

The same construction presented above can be repeated for systems obeying statistics other than microcanonical
\cite{Campisi07PHYSA385,Campisi09PRE80}.
If applied to the canonical ensemble,
 $\rho(\bm{\xi};\bm{\lambda},\beta) = e^{-\beta H(\bm{\xi};\bm{\lambda})}/Z(\bm{\lambda},\beta)$, (with $Z(\bm{\lambda},\beta)=\Tr \, e^{-\beta H(\bm{\xi};\bm{\lambda})}$ being the canonical partition function)
the canonical expression
\begin{align}
F_i(E,\bm{\lambda}) = -\Tr\, \left( \frac{\partial H}{\partial \lambda_i} \frac{ e^{-\beta H(\bm{\xi};\bm{\lambda})}}{Z(\bm{\lambda},\beta)}  \right ) 
\end{align}
for the forces, along with the equation of state of the ideal gas, \emph{uniquely} identifies the canonical parameter $\beta$ as the integrating factor, and its associated primitive
\begin{align}
S(\beta,\bm{\lambda})= -k_B \beta^2 \frac{\partial }{\partial \beta} \frac{\ln Z(\bm{\lambda},\beta)}{\beta} 
\end{align}
as the only thermodynamically consistent expressions of inverse temperature and entropy within the canonical formalism.\footnote{Incidentally  $S(\beta,\bm{\lambda})= -k_B \Tr \rho(\bm{\xi};\bm{\lambda},\beta) \ln \rho(\bm{\xi};\bm{\lambda},\beta)$, that is, the canonical entropy coincides with the Gibbs-von Neumann information of the canonical distribution $ \rho(\bm{\xi};\bm{\lambda},\beta) $.}
In the canonical formalism nothing formally constraints the sign of $\beta$ to be definite. A spin system in a canonical state at negative $\beta$ will have a positive internal energy $U$. The same system 
in the microcanonical state of energy $E=U$ will, however, have a positive thermodynamic temperature. This evidences the inequivalence of canonical and
microcanonical ensembles in systems with a finite spectrum.

\subsection{Exact vs approximate constructions}
In an attempt to justify the correctness of the Boltzmann entropy, Frenkel and Warren \cite{Frenkel15AJP83}, provided a construction which leads to the Boltzmann entropy. It must be stressed that the construction presented by Frenkel and Warren \cite{Frenkel15AJP83} is approximate, and valid only under the assumption that the saddle point approximation holds. This approximation holds only when the density of states increases exponentially with the energy. Under this assumption, however, the density of states $\omega$ and phase volume $\Omega$ coincide. So the construction of Frenkel and Warren \cite{Frenkel15AJP83} cannot shed light onto which entropy expression is appropriate in the case when they do not coincide, which is indeed the very case of practical interest.

In contrast, the present construction is exact, i.e., it holds regardless of the functional dependence of the density of states on energy. Accordingly it says that in any case, independent of whether equivalence of the two entropies holds, the volume entropy is the consistent choice.

\subsection{Thermodynamic temperature equals equipartition temperature}
For continuous classical Hamiltonian systems, thanks to the equipartition theorem \cite{Khinchin49Book}, the thermodynamic temperature $T$ is identical with the equipartition temperature $T_\text{eq}$:
\begin{align}
k_B T_\text{eq}\doteq \left \langle \xi_k\frac{\partial H}{\partial \xi_k}\right \rangle  = \frac{\Omega}{\omega}= k_B (\partial_E S_G)^{-1}=k_B T\, .\label{eq:equipartiton}
\end{align}
where the average is the microcanonical average on the shell $(E,\bm{\lambda})$.
This provides further evidence that the choice $g(x)=k_B \ln x$ conforms to the common notion of temperature of any classical system, not just the ideal monoatomic gas. We further remark that the equipartition theorem also identifies the temperature $T(E,\bm{\lambda})$ in Eq.  (\ref{eq:PartialLogPhi}) as an intensive quantity, namely a property that is equally shared by all subsystems \cite{Hilbert14PRE90}.

We emphasize that at variance with previous approaches to the foundations of the Gibbs entropy \cite{Hertz10AP338a,Berdichevsky97Book,Campisi05SHPMP36,Campisi10AJP78}, which postulated that the thermodynamic temperature is the equipartition temperature,  here we have instead postulated only that temperature is the integrating factor that is consistent with the ideal gas law and have obtained the coincidence with the equipartition temperature as an aftermath. 
The advantage of the present approach is evident: it applies to any microcanonical system, even those for which there is no equipartition theorem (e.g., quantum systems).

\subsection{the Boltzmann entropy fails to predict the value of thermodynamic forces}
\label{subsec:failure}
At variance with other approaches we chose as starting point the expression for the microcanonical forces (\ref{eq:Fmicro}) which is universally agreed upon and built our construction on that firm ground. The salient point of our argument is the identity (\ref{eq:Fomega}) expressing the microcanonical forces in terms of the partial derivatives of $\Omega$. The identity (\ref{eq:Fomega}) alone has as a consequence that the entropy must be of the form $S_g=g(\Omega)$ with some $g$ with non null derivative $g'$. In fact, for any $S_g$ one finds the forces $F_i^g= \partial_i S_g/\partial_E S_g$, to be identical to the microcanonical forces $F_i$, Eq. (\ref{eq:Fmicro})
\begin{align}
F_i^g = \frac{\partial_i S_g}{\partial_E S_g}=  \frac{g'\partial_i \Omega}{g' \partial_E \Omega}= \frac{\partial_i \Omega}{\partial_E \Omega} = F_i \, .
\end{align}
Here $\partial_i$ is a shorthand notation for $\partial/\partial \lambda_i$.
If one employs an entropy expression that is not of the form $g(\Omega)$, e.g., the Boltzmann entropy, one can well end up in wrongly evaluating the forces. 

This happens, for example, in the case of a large collection of $N \gg 1$ non interacting $1/2$ spins in a magnetic field $B$, at energy $E$ \cite{Dunkel14NATPHYS10}, that is the prototypical example of the emergence of negative Boltzmann temperature \cite{Purcell51PR81,Ramsey56PR103}.
The Hamiltonian reads \cite{Kubo65book}
\begin{align}
H =- B  \mu \sum_{i=1}^N \sigma^i \, .
\end{align}
Here $B$ plays the role of the external parameter  $\lambda$, $\sigma^i$ is $\pm 1$ depending on whether the spin points
parallel (up) or antiparallel (down) to the field, and $\mu$ is the magnetic moment of each spin.
At energy $E$, the magnetization is given by (\ref{eq:Fmicro}):
\begin{align}
M(E,B) = -\langle \partial_B H\rangle = - \langle H\rangle/B =- E/B \, .
\label{eq:exact}
\end{align}
The number of states with $n$ spins up is
\begin{align}
\mathcal W_\omega(n) = \frac{N!}{n!(N-n)!}\, .
\end{align}
The number of states with no more than $n$ spins up  is
\begin{align}
\mathcal W_\Omega(n) =\sum_{k=0}^n \frac{N!}{k!(N-k)!} \, .
\end{align}
Using the relation $E=-(2n-N)\mu B$, and treating $E$ as a continuous variable under the assumption that $N$ is very large, according to standard 
procedures, we observe that $\mathcal W_\omega(N/2-E/2\mu B)$ denotes the number of states with energy between $E-\mu B$ and $E+ \mu B$.
The density of states is therefore:
\begin{align}
\omega(E,B) &= \frac{\mathcal W_\omega (N/2-E/2\mu B)}{2\mu |B|}\, ,
\label{eq:omega}
\end{align}
and the number of states with energy below $E$ is 
\begin{align}
\Omega(E,B) &= \mathcal W_\Omega(N/2-E/2\mu B) \, .
\label{eq:Omega}
\end{align}
\begin{figure}[]
		\includegraphics[width=.4\textwidth]{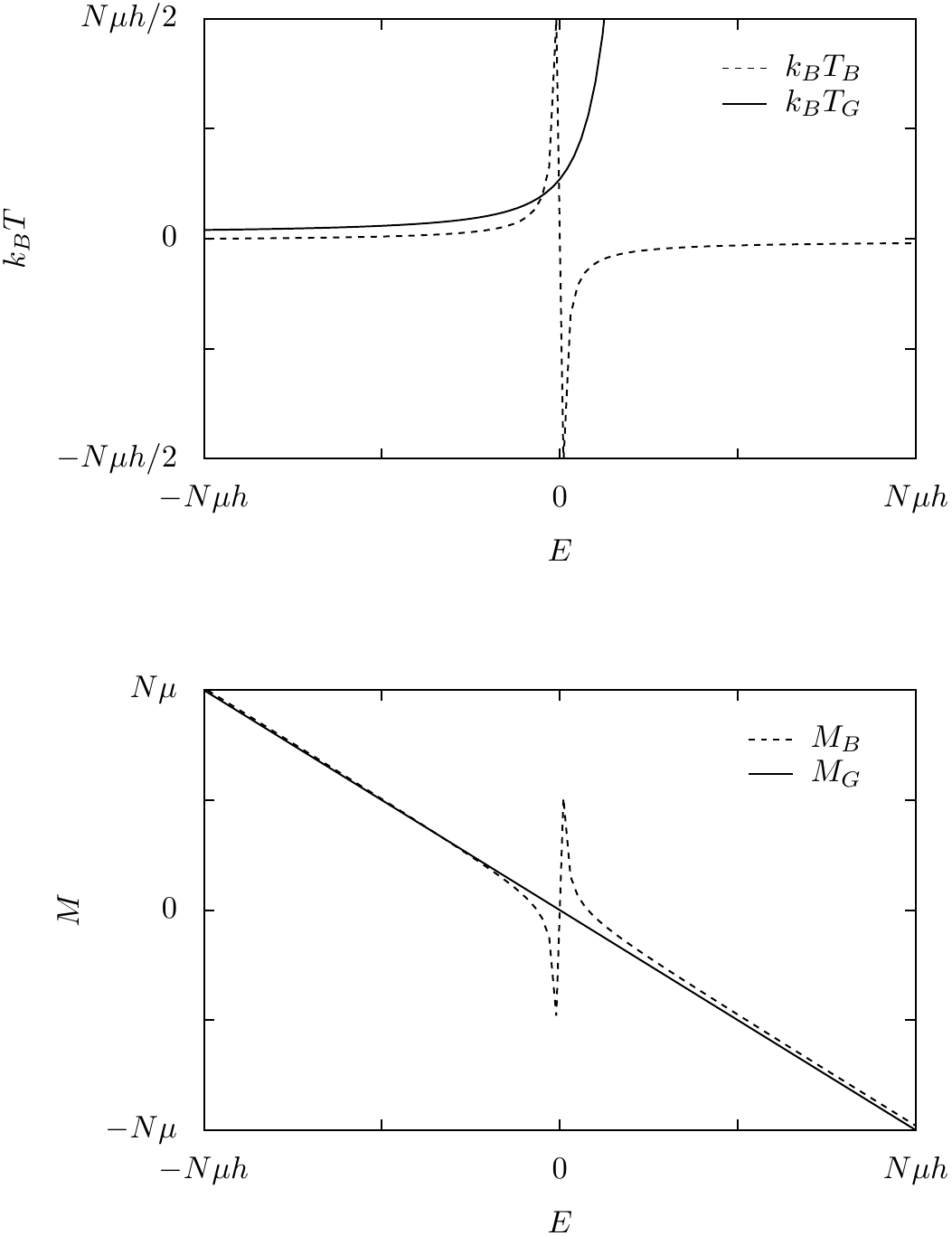}
		\caption{Temperature $T$ and magnetization $M$ of a system of $N$ non-interacting $1/2$ spins, as predicted by the Boltzmann entropy $S_B$, and the Gibbs entropy $S_G$ Here $N=100$. Only Gibbs magnetization conforms with the physical magnetization $M=-E/B$.}
		\label{fig:Fig1}
\end{figure}
Figure \ref{fig:Fig1} shows the Gibbs and Boltzmann temperatures and magnetizations as functions of $E$ calculated with
\begin{align}
k_B T_B &= \frac{k_B}{\partial_E S_B}=\frac{\omega}{\partial_E \omega};\quad
M_B = \frac{\partial_B S_B}{\partial_E S_B} = \frac{\partial_B \omega}{\partial_E \omega} \label{eq:TMB}\, ,\\
k_B T_G &= \frac{k_B}{\partial_E S_G}=\frac{\Omega}{\partial_E \Omega};\quad
M_G = \frac{\partial_B S_G}{\partial_E S_G} = \frac{\partial_B \Omega}{\partial_E \Omega}\, . \label{eq:TMG}
\end{align}
For larger values of $N$ qualitatively similar plots are obtained. 
A very unphysical property of $T_B$ is that with the flip of a single spin
it jumps discontinuously from $+\infty$ to $-\infty$ in the thermodynamic limit. The usual reply to such a criticism would be, following \cite{Ramsey56PR103}, to say that one should look instead at the quantity $-1/T_B$, which displays no divergence. 
No way out is however possible if one considers the magnetization.
As can be seen from the figure, only $S_G$ reproduces the exact result, Eq. (\ref{eq:exact}) whereas the magnetization given by $S_B$ is drastically off, and even predicts a nonexistent and unphysical phase transition, in the thermodynamic limit, where the magnetization abruptly jumps from $-\infty$ to $+\infty$ as a single spin flips from $+1$ to $-1$.
The results in the figure are also corroborated by analytical calculations. Using Eqs. (\ref{eq:TMB}) and (\ref{eq:TMG}) with Eqs. (\ref{eq:omega}) and (\ref{eq:Omega}) we obtain
\begin{align}
M_B &= -(E + k_B T_B)/B \label{eq:MB} \, ,\\ 
M_G & = -E/B = M\, .
\end{align}
Thus the discrepancy $\Delta$ between the Boltzmann magnetization and the physical magnetization is given by the negative Boltzmann thermal energy rescaled by the applied magnetic field:
\begin{align}
\Delta \doteq M_B-M=  -k_B T_B/B\, .
\end{align}
Since $T_B$ diverges around the zero energy in the thermodynamic limit, so does the discrepancy $\Delta$. Note that the discrepancy also diverges as the intensity of the applied magnetic field decreases. It is interesting to notice that, while in the thermodynamic limit $T_G$ approaches $T_B$ for $E<0$, the same is not true for $M_B$, which distinctly deviates from $M=M_G$ for both $E>0$ and $E<0$. This unveils the fact, apparently previously unnoticed, that Boltzmann and the Gibbs entropy are not equivalent even in the lower part of the spectrum of large spin systems.
\begin{figure}[]
		\includegraphics[width=.4\textwidth]{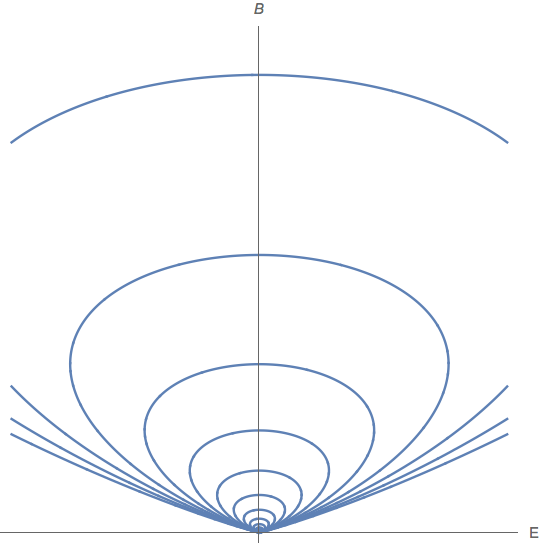}
		\caption{Iso-$S_B$ lines do not coincide with the adiabats $E/B=\text{const}$. }
		\label{fig:Fig2}
\end{figure}

Equation (\ref{eq:MB}) is a special case of a general relation linking the Boltzmann forces ($F_B^i=\partial_i S_B/\partial_E S_B$) and the Gibbs forces $F_B^i$ (i.e., the thermodynamic forces $F^i$), reading:
\begin{align}
F_B^i-F^i = k_B T_B \frac{ \partial F^i}{\partial E}\, . \label{eq:MBvsMG}
\end{align}
This equation accompanies a similar relation linking Boltzmann and Gibbs temperatures
\begin{align}
T_B= \frac{T_G}{1- k_B C_G^{-1}}\, . \label{eq:TBvsTG}
\end{align}
with $C_G=(\partial_E T_G)^{-1}$ being the heat capacity. Equations (\ref{eq:MBvsMG}) and (\ref{eq:TBvsTG}) follow by taking the derivative with respect to $E$ of $F_i$ and $T_G$ respectively.

The reason for the thermodynamic inconsistency of $S_B$ (consistency of $S_G$) can also be understood in the following way. Consider the heat differential $\delta Q = dE + MdB= dE-(E/B) dB$. Clearly $1/E$ is an integrating factor: $\delta Q/E = dE/E - dB/B=d\ln(E/B)$. Hence $f(E,B)=\ln (E/B)$ is a primitive. Accordingly the adiabats are determined by the equation
\begin{align}
E/B = \text{const} \qquad \text{(adiabats equation)}\, ,
\label{eq:adiabats}
\end{align}
and the entropy must be some monotonic function of $\ln E/B$, that is of $E/B$. By inspecting Eqs. (\ref{eq:omega}) and (\ref{eq:Omega}) we see that the phase volume $\Omega$ is a monotonic function of $E/B$ while the density of states $\omega$ is not a function of $E/B$; hence $S_B$ is thermodynamically inconsistent.

The inequivalence of $S_G$ and $S_B$ is most clearly seen by plotting the iso-$S_B$ lines in the thermodynamic space $E,B$; see Fig. \ref{fig:Fig2}. Note that the adiabats, Eq. (\ref{eq:adiabats}) are straight lines passing through the origin. The iso-$S_B$ lines instead predict a completely different structure of the adiabats. Note in particular that the iso-$S_B$ lines are closed. This evidences their thermodynamical inconsistency. 

Summing up: the Boltzmann entropy severely fails to accomplish one of its basic tasks, namely, reproducing the correct value of the thermodynamic forces and of heat.

\section{Concluding remarks}
We have shown that, within the microcanonical formalism there is only one possible choice of entropy that is consistent with the second law and the equation of state of an ideal gas, namely, the Gibbs entropy. Discarding the Gibbs entropy in favour of the Boltzmann entropy, may accordingly result in inconsistency with either of those two pillars. For the great majority of large thermodynamic systems, Gibbs and Boltzmann entropies practically coincide; hence there is no problem regarding which we choose. However, there are cases when the two do not coincide: examples are spin systems  
\cite{Dunkel14NATPHYS10} and point vortex gases \cite{Berdichevsky95PRE51}, where Boltzmann temperature, in disagreement with Gibbs temperature, has no definite sign, and the Boltzmann entropy can largely fail to predict correct values of thermodynamic forces. 

It must be stressed that the demonstrated failure of the Boltzmann entropy to reproduce the thermodynamic forces is not  restricted to small systems, where the failure was already known to occur \cite{Dunkel14NATPHYS10}, but survives, and even becomes more prominent, in the thermodynamic limit, where the Boltzmann entropy predicts an unphysical and nonexistent phase transition in the magnetization of a system of non-interacting spins in a magnetic field.

In the light of the present results, together with the established fact that the Gibbs entropy conforms with all thermodynamic laws \cite{Hilbert14PRE90}, the issue of which entropy expression is correct is apparently now fully and ultimately settled.

\subsection*{Acknowledgements}
The author is indebted to J\"orn Dunkel, Stefan Hilbert, Peter Talkner and especially Peter H\"anggi, for the many discussions we had on this topic for years. This research was supported by a Marie Curie Intra European Fellowship within the 7th European Community Framework Programme through the project NeQuFlux Grant No. 623085 and by the COST Action No. MP1209 ``Thermodynamics in the quantum regime.''

\end{document}